\documentclass[english]{llncs}
\usepackage[T1]{fontenc}
\usepackage[latin9]{inputenc}
\usepackage{color}
\usepackage{amsmath}
\usepackage{amssymb}
\usepackage{graphicx}
\usepackage{esint}
\usepackage{epstopdf}
\usepackage{epsfig}
\usepackage{cite}
\usepackage{babel}
\begin{document}

\title{General structures of reversible and quantum gates}

\author{Kishore Thapliyal\thanks{Email:tkishore36@yahoo.com}, Anirban Pathak\thanks{Email:anirban.pathak@gmail.com}}

\institute{Jaypee Institute of Information Technology, A-10, Sector-62, Noida,
India}
\maketitle
\begin{abstract}
The most general structure (in matrix form) of a single-qubit gate
is presented. Subsequently, used that to obtain a set of conditions
for testing (a) whether a given 2-qubit gate is genuinely a 2-qubit
gate, i.e., not decomposable into two single qubit gates and (b) whether
a given single qubit gate is self-inverse? Relevance of the results
reported here is discussed in the context of optimization of reversible
and quantum circuits, especially for the optimization of quantum cost.
A systematic procedure is developed for the identification of the
non-decomposable 2-qubit gates. Such a non-decomposable 2-qubit gate
along with all possible single qubit gates form a universal quantum
gate library. Further, some possible applications of the present work
are also discussed.
\end{abstract}

\section{Introduction}

The use of quantum resources provides an enhancement in the performance
of certain tasks in comparison to their classical counterparts. To
be specific, a quantum computer can perform factorization \cite{shor}
and unsorted database search \cite{grover} with a speed not achievable
by its classical counterpart, and quantum cryptography can provide
unconditional security \cite{book}, a desirable feature of secure
communication that cannot be achieved by any classical protocol. These
facts lead to a simple question, how are these tasks performed in
the quantum world? A simple answer would be by exploiting quantum
superposition through suitable quantum operations which may be viewed
as quantum gates. It is also worth noting that all the operations,
besides measurement and noise, are unitary in nature. Therefore, quantum
operations (except quantum measurement and Kraus operators representing
various noise models) are essentially reversible in nature. In what
follows, we refer to these unitary quantum operators that actually
describe evolution of a quantum state and map the initial states to
final states as quantum gates and analyze their properties.

Before, we proceed further, we would like to note that all quantum
gates are reversible, but they are usually referred to as quantum
gates, whereas by a reversible gate we usually mean a classical reversible
gate which is also described by a unitary operation ($U)$. This is
so because unitarity ensures reversibility through the condition $U^{\dagger}=U^{-1}$
or in other words, unitarity ensures the existence of $U^{-1}$ for
every unitary operation $U$ and thus establishes reversibility. Consequently,
the basic structure of classical reversible and quantum gates are
the same, In fact, ${\rm CNOT,}\,{\rm SWAP}$ and many other gates
work in both classical and quantum domain, they are described by the
same matrices, but there is a small difference as far as the acceptable
input states are concerned. A classical reversible gate cannot accept
superposition states as input, whereas a quantum gate can. Specially,
a reversible gate should not accept a superposition state as input
at the controlled bit. For example, if a ${\rm CNOT}$ gate accepts
a superposition state in the controlled bit/qubit then the output
will be an entangled state, which has no classical analogue and which
cannot exist in the classical world. Thus, a classical reversible
gate and a quantum gate would have the same mathematical form (both
being described by unitary matrices), whether the quantum gate will
have a classical counterpart or not would depend on the input-output
relation; specifically, on whether the gate produces classically acceptable
outputs for valid classical inputs. As quantum gates are more general
and classical reversible gates form only a subclass of them, in what
follows our discussion will be focused on the quantum gates (unless
otherwise stated) only, and we will mostly focus on the structure
of the unitary operators (gates) without providing much attention
to whether a particular gate has a classical counterpart or not. However,
the analysis is valid in general for both quantum and reversible gates
and may be useful in optimizing both quantum and reversible circuits.

We have already noted that quantum gates are more general and classical
reversible gates are special cases of them. However, the notion of
quantum computer in general and quantum Turing machine in particular
\cite{QuTur} has originated from the idea of reversible Turing machine
\cite{Bennett}, which was proposed to perform computation in reversible
manner so that heat loss due to erasure of information predicted by
Landauer's principle \cite{Landauer61} can be circumvented. Heat
loss is a major issue in today's VLSI technology. This is so because
in accordance with Moore's law \cite{moore}, the number of transistors
per unit area is doubling in every 18 months, and proportionately
length of interconnecting wires and energy losses through those wires
are also increasing. In case, we can make a room temperature superconducting
material, we will be able to avoid the $I^{2}R$ type of loss that
happens through these wires. However, in the traditional irreversible
computing, some losses of energy would still happen as it is implemented
with irreversible logic gates, like AND, NOR, NAND all of which map
a 2-bit input into a 1-bit output and thus causes an energy loss amounting
to at least $kT\ln2$. This advantage of reversible computing and
the computational speed up achieved by the quantum computer have motivated
scientists to design and optimize reversible and quantum circuits
for various purposes {[}for a set of interesting reversible circuits
see \cite{rev-comp,Revlib}, some interesting reversible circuits
and their optimization are reported in \cite{rev1,rev2,rev3}, whereas
a set of important quantum circuits can be found at \cite{ECP,QC1,QC2,QC3,QC4,QC5,QC6}{]}.
All these circuits are designed using gates represented by unitary
operations, but the general structure of those unitary operations
is not investigated until today. Motivated by this fact here we report
some observations on the general structures of the unitary gates and
also discuss how to exploit those observations to perform optimization
of the reversible/quantum circuits. To be precise, we would first
describe the general structure of single qubit gates and using that
we will obtain a set of requirements to be satisfied by a 2-qubit
decomposable gate (i.e., a 2-qubit gate which can be expressed as
tensor product of two single qubit gates). Further, we would also
formulate a method to obtain those two single qubit gates once it
is ensured that the given 2-qubit gate is decomposable. Finally, some
applications of the decomposability tests designed in this paper are
discussed with specific attention toward a physical problem involving
different type of beam-splitters and the problems related to optimization
of various types of cost metrics (such as gate count or circuit complexity,
depth and width of a circuit, quantum cost) associated with quantum
circuits.

The rest of the paper is organized as follows. In Sec. \ref{sec:General-structure-for},
we derive a general structure of single qubit gates. Using this general
structure, in Sec. \ref{sec:Hermiticity-or-self}, we have obtained
a general form to be possessed by the single qubit Hermitian unitary
operations, which would represent self-inverse quantum gates. We further
use the general structure to formulate the conditions to check whether
a given 2-qubit gate is decomposable or not in Sec. \ref{sec:Decomposability-of-a}.
In Sec. \ref{sec:Two-single-qubit}, a method to reconstruct the decomposed
single qubit operations from the given decomposable 2-qubit gate is
designed. Finally, the paper is concluded in Sec. \ref{sec:Conclusion}
after discussing some applications of the present results in Sec.
\ref{sec:Applications}.

\section{General structure for single qubit gates\label{sec:General-structure-for}}

As we mentioned in the previous section, we are interested in obtaining
the general structure of the unitary gates. To begin with, let us
consider an arbitrary single qubit gate, 
\begin{equation}
U=\left[\begin{array}{cc}
a & b\\
c & d
\end{array}\right],\label{eq:unitary-mat}
\end{equation}
which would satisfy $U^{\dagger}U=UU^{\dagger}=\mathbb{I},$ being
unitary. Here, the first condition (i.e., $U^{\dagger}U=\mathbb{I})$
gives us 
\begin{equation}
U^{\dagger}U=\left[\begin{array}{cc}
\left|a\right|^{2}+\left|c\right|^{2} & a^{*}b+c^{*}d\\
ab^{*}+cd^{*} & \left|b\right|^{2}+\left|d\right|^{2}
\end{array}\right]=\mathbb{I},\label{eq:unitarity1}
\end{equation}
while the second one (i.e., $UU^{\dagger}=\mathbb{I}$) yields 
\begin{equation}
UU^{\dagger}=\left[\begin{array}{cc}
\left|a\right|^{2}+\left|b\right|^{2} & ac^{*}+bd^{*}\\
a^{*}c+b^{*}d & \left|c\right|^{2}+\left|d\right|^{2}
\end{array}\right]=\mathbb{I}.\label{eq:unitarity2}
\end{equation}
From Eq. (\ref{eq:unitarity1}) we can easily observe that \begin{subequations}

\begin{equation}
\left|a\right|^{2}+\left|c\right|^{2}=1,\label{eq:FromUni1a}
\end{equation}
\begin{equation}
\left|b\right|^{2}+\left|d\right|^{2}=1,\label{eq:FromUni1b}
\end{equation}
and 
\begin{equation}
ab^{*}+cd^{*}=0.\label{eq:FromUni1c}
\end{equation}
\end{subequations}Similarly, from Eq. (\ref{eq:unitarity2}) we obtain\begin{subequations}

\begin{equation}
\left|a\right|^{2}+\left|b\right|^{2}=1,\label{eq:FromUni2a}
\end{equation}
\begin{equation}
\left|c\right|^{2}+\left|d\right|^{2}=1,\label{eq:FromUni2b}
\end{equation}
and 
\begin{equation}
ac^{*}+bd^{*}=0.\label{eq:FromUni2c}
\end{equation}
\end{subequations}In Eq. (\ref{eq:FromUni2a}), we can substitute
$\left|a\right|=\cos\theta=\sqrt{1-\left|b\right|^{2}}.$ This substitution
and the use of Eqs. (\ref{eq:FromUni1a}) and (\ref{eq:FromUni2a})
would yield $\left|c\right|=\left|b\right|=\sin\theta.$ Similarly,
using Eq. (\ref{eq:FromUni1b}) in Eq. (\ref{eq:FromUni2a}), we obtain
$\left|d\right|=\left|a\right|=\cos\theta.$ Using these values, we
can rewrite the arbitrary single qubit unitary matrix $U$ described
by (\ref{eq:unitary-mat}) as 
\begin{equation}
U=\left[\begin{array}{cc}
\cos\theta\exp\left(i\phi_{11}\right) & \sin\theta\exp\left(i\phi_{12}\right)\\
\sin\theta\exp\left(i\phi_{21}\right) & \cos\theta\exp\left(i\phi_{22}\right)
\end{array}\right],\label{eq:NewU}
\end{equation}
where we have used the polar form of complex elements of the unitary
matrix in Eq. (\ref{eq:unitary-mat}). Further, using Eqs. (\ref{eq:FromUni1c})
and (\ref{eq:FromUni2c}) with values of different elements of matrix
$U$ as given in Eq. (\ref{eq:NewU}), we obtain the same condition
from both the equations, i.e., 
\[
\exp\left(i\phi_{12}-i\phi_{22}\right)=-\exp\left(i\phi_{11}-i\phi_{21}\right).
\]
From which we can easily write 
\begin{equation}
\phi_{12}-\phi_{22}=\phi_{11}-\phi_{21}\pm\left(2k+1\right)\pi,\label{eq:phase-condition}
\end{equation}
with $k$ being an integer. We may consider an angle $\phi_{0}$ in
such a way that $a$ and $d$ in Eq. (\ref{eq:unitary-mat}) can be
written as complex conjugates of each other, which means 
\begin{equation}
\phi_{11}-\phi_{0}=\phi_{0}-\phi_{22}\label{eq:globalphase}
\end{equation}
or 
\[
\phi_{0}=\frac{1}{2}\left(\phi_{11}+\phi_{22}\right).
\]
Interestingly, the same angle $\phi_{0}$ also makes $b$ and $c$
complex conjugates of each other as 
\begin{equation}
\phi_{21}-\phi_{0}=\phi_{0}-\phi_{12}\pm\left(2k+1\right)\pi.\label{eq:phase-condition-2}
\end{equation}
In fact, this equation can also be obtained by using Eq. (\ref{eq:globalphase})
in Eq. (\ref{eq:phase-condition}). Finally, we may rewrite $U$ in
Eq. (\ref{eq:NewU}) as 
\begin{equation}
U=\exp\left(i\phi_{0}\right)\left[\begin{array}{cc}
\cos\theta\exp\left(i\phi_{1}\right) & \sin\theta\exp\left(i\phi_{2}\right)\\
-\sin\theta\exp\left(-i\phi_{2}\right) & \cos\theta\exp\left(-i\phi_{1}\right)
\end{array}\right],\label{eq:finalU}
\end{equation}
where 
\[
\phi_{0}=\frac{1}{2}\left(\phi_{11}+\phi_{22}\right),
\]
\[
\phi_{1}=\frac{1}{2}\left(\phi_{11}-\phi_{22}\right),
\]
and 
\[
\phi_{2}=\frac{1}{2}\left(\phi_{12}-\phi_{21}\mp\left(2k+1\right)\pi\right).
\]

It is noteworthy here that four equivalent general structures of single
qubit unitary operation can be written by changing the position of
negative sign among the four elements of the matrix in Eq. (\ref{eq:finalU}).
Actually unitarity demands that one of the elements (expressed in
the polar form) of the single qubit unitary operator has to have a
sign opposite to that of the other three elements (say, negative sign,
when the rest of the elements are with positive sign). Thus, the negative
sign put in front of any of the 4 elements of the above structure
of the unitary operator, would also provide a general structure of
the single qubit unitary operator.

\section{Hermiticity or self reversibility of single qubit gates\label{sec:Hermiticity-or-self}}

Hermitian matrices are the one satisfying $A^{\dagger}=A$. In case
of unitary matrices, it turns out to be $A^{\dagger}=A^{-1}=A$. Thus,
if we find that a unitary matrix that represents a gate is also Hermitian,
then we would be able to conclude that the gate is self inverse, too
\cite{non}. For obtaining the general structure of self-reversible
unitary operations, if we write the adjoint of $U$ in Eq. (\ref{eq:finalU})
as 
\begin{equation}
U^{\dagger}=\exp\left(-i\phi_{0}\right)\left[\begin{array}{cc}
\cos\theta\exp\left(-i\phi_{1}\right) & -\sin\theta\exp\left(i\phi_{2}\right)\\
\sin\theta\exp\left(-i\phi_{2}\right) & \cos\theta\exp\left(i\phi_{1}\right)
\end{array}\right],\label{eq:udagger}
\end{equation}
then we can see that Hermiticity condition $U^{\dagger}=U$, would
yield 
\[
\exp\left(i\phi_{0}+i\phi_{1}\right)=\exp\left(-i\phi_{0}-i\phi_{1}\right),
\]
\[
\exp\left(i\phi_{0}+i\phi_{2}\right)=-\exp\left(-i\phi_{0}+i\phi_{2}\right),
\]
 and
\[
\exp\left(i\phi_{0}-i\phi_{1}\right)=\exp\left(-i\phi_{0}+i\phi_{1}\right).
\]
By solving this set of conditions on the phase parameters, one can
obtain $\phi_{0}=\left(\frac{2k+1}{2}\right)\pi$ and $\phi_{1}=-\frac{\pi}{2}$.
Hence, the mathematical structure of a Hermitian unitary matrix (Hermitian
single qubit quantum gate) is
\[
U_{H}=\pm\left[\begin{array}{cc}
\cos\theta & i\sin\theta\exp\left(i\phi_{2}\right)\\
-i\sin\theta\exp\left(-i\phi_{2}\right) & -\cos\theta
\end{array}\right].
\]

It should be noted here that permutation of columns of unitary (\ref{eq:finalU})
will not change the mathematical structure of Hermitian quantum gate
obtained here as the constraint equation will remain unchanged after
such a permutation \cite{non}. For convenience, we may describe this
gate as $U_{H}\left(\theta,\phi_{2}\right)$ and note that the well
known single qubit gates can be expressed in this notation as $H=U_{H}\left(\frac{\pi}{4},\frac{3\pi}{2}\right),\,X=U_{H}\left(\frac{\pi}{2},\frac{3\pi}{2}\right),\,iY=U_{H}\left(\frac{\pi}{2},0\right),$
and $Z=U_{H}\left(0,\phi_{2}\right),$ whereas a phase gate $P=\left[\begin{array}{cc}
1 & 0\\
0 & \exp\left(i\xi\right)
\end{array}\right]\forall\xi\ne n\pi$ is not self inverse and cannot be expressed in the above form.

\section{Decomposability of a two qubit gate using the general structure of
single qubit gates\label{sec:Decomposability-of-a}}

In this section, we wish to formulate a method to distinguish between
a set of 2-qubit gates decomposable in two single qubit gates from
the genuine 2-qubit gates or non-decomposable gates.  For the same,
we have used $U$ obtained in the last section, given in Eq. (\ref{eq:finalU}),
as the most general form of single qubit gates and checked the decomposability
of two qubit gates. To do so, let us consider two arbitrary single
qubit gates $U_{1}$ and $U_{2}$ having the following form\begin{subequations}
\begin{equation}
U_{1}=\exp\left(i\phi_{1}\right)\left[\begin{array}{cc}
u_{1} & u_{2}\\
-u_{2}^{*} & u_{1}^{*}
\end{array}\right]\label{eq:U1-U2-1}
\end{equation}
 and 
\begin{equation}
U_{2}=\exp\left(i\phi_{2}\right)\left[\begin{array}{cc}
v_{1} & v_{2}\\
-v_{2}^{*} & v_{1}^{*}
\end{array}\right].\label{eq:U1-U2-2}
\end{equation}
\end{subequations}It is easy to write the tensor product of these
two single qubit gates as 
\begin{equation}
U=U_{1}\otimes U_{2}=\exp\left(i\phi_{1}+i\phi_{2}\right)\left[\begin{array}{cccc}
u_{1}v_{1} & u_{1}v_{2} & u_{2}v_{1} & u_{2}v_{2}\\
-u_{1}v_{2}^{*} & u_{1}v_{1}^{*} & -u_{2}v_{2}^{*} & u_{2}v_{1}^{*}\\
-u_{2}^{*}v_{1} & -u_{2}^{*}v_{2} & u_{1}^{*}v_{1} & u_{1}^{*}v_{2}\\
u_{2}^{*}v_{2}^{*} & -u_{2}^{*}v_{1}^{*} & -u_{1}^{*}v_{2}^{*} & u_{1}^{*}v_{1}^{*}
\end{array}\right].\label{eq:u1tensoru2}
\end{equation}
We can also consider an arbitrary $4\times4$ matrix (which is assumed
to represent a 2-qubit gate) as 
\begin{equation}
A=\exp\left(i\phi\right)\left[\begin{array}{cccc}
a & b & c & d\\
e & f & g & h\\
i & j & k & l\\
m & n & o & p
\end{array}\right]\label{eq:arbitrary2qubit}
\end{equation}
and compare the coefficients of $A$ with the matrix $U$ in Eq. (\ref{eq:u1tensoru2})
to find the condition of separability (decomposability), the satisfaction
of which for a given 2-qubit gate would mean that the given two qubit
gate is not genuinely a 2-qubit gate as it can be decomposed into
two single qubit gates. By comparing the elements of (\ref{eq:u1tensoru2})
and (\ref{eq:arbitrary2qubit}) we can observe the following conditions:
\begin{description}
\item [{Test~1:}] $p=a^{*},$ $o=-b^{*},$ $n=-c^{*},$ and $m=d^{*}.$
\item [{Test~2:}] $f=k^{*}=a\exp\left(i\phi_{a}\right).$
\item [{Test~3:}] $e=-l^{*}=b\exp\left(i\phi_{b}\right).$
\item [{Test~4:}] $h=-i^{*}=c\exp\left(i\phi_{c}\right).$
\item [{Test~5:}] $g=j^{*}=d\exp\left(i\phi_{d}\right).$
\end{description}
Failure of any of these tests would imply the inseparability of the
2-qubit gate under consideration, which would mean that the investigated
2-qubit gate is genuinely a 2-qubit gate, i.e., the 2-qubit gate cannot
be decomposed into two single qubit gates operating on each qubit,
and such a 2-qubit gate may be used to construct a universal gate library
in association with all the single qubit gates. This is so because
it is well known that any genuine two qubit gate and set of all single
qubit gates form a universal quantum gate library \cite{DiVincenzo}. 

Note that to test the decomposability of a given 2-qubit gate, the
2-qubit gate is to be written in the form of matrix $A$ in Eq. (\ref{eq:arbitrary2qubit}),
where to obtain $\phi$ we can again use the same method as was used
in Eq. (\ref{eq:globalphase}). Specifically, we may take the global
phase $\phi$ in such a way that one of the conditions in Tests 1-5
is satisfied. In case of a decomposable 2-qubit gate all the remaining
conditions should also remain valid. Without loss of generality, considering
the first condition in Test 1, we can obtain $\exp\left(i\phi\right)=\frac{\left(\frac{a_{11}}{\left|a_{11}\right|}+\frac{a_{44}}{\left|a_{44}\right|}\right)}{\left|\frac{a_{11}}{\left|a_{11}\right|}+\frac{a_{44}}{\left|a_{44}\right|}\right|},$
where $a_{11}=a\exp\left(i\phi\right)$ and $a_{44}=p\exp\left(i\phi\right)$
are the first and sixteenth elements of the arbitrary 2-qubit unitary
before taking a common phase out of the matrix $\left[a_{ij}\right]$.
This global phase will be equivalent to $\exp\left(\phi_{1}+\phi_{2}\right)$
in Eq. (\ref{eq:u1tensoru2}). 

The present study also reveals that to check the inseparability of
a given 2-qubit gate one can perform an assessment of the unitary
before writing the unitary in the form of matrix $A$ given in Eq.
(\ref{eq:arbitrary2qubit}). Specifically, all the diagonal elements
in a decomposable 2-qubit unitary have the same modulus. Similarly,
all the anti-diagonal elements in a decomposable 2-qubit unitary also
have the same modulus, which is independent of the value for the diagonal
elements.

Therefore, we may now summarize the conditions that ensure that a
2-qubit gate is decomposable into two single qubit gates as follows:
\begin{description}
\item [{Condition~1:}] All the diagonal elements have the same modulus
value. Even the anti-diagonal elements also have the same modulus,
but not necessarily equal to the main diagonal elements. In other
words, $\left|a_{11}\right|=\left|a_{22}\right|=\left|a_{33}\right|=\left|a_{44}\right|$
and $\left|a_{41}\right|=\left|a_{32}\right|=\left|a_{23}\right|=\left|a_{14}\right|$,
where $a_{ij}$ are the elements of the given matrix before writing
in the form of matrix $A$ of Eq. (\ref{eq:arbitrary2qubit}).
\item [{Condition~2:}] After writing the given matrix in the form of matrix
$A$ of Eq. (\ref{eq:arbitrary2qubit}) all Tests 1-5 are satisfied.
\end{description}
Here, it is important to note that fulfillment of Condition 2 ensures
that Condition 1 is also satisfied but not vice versa. Therefore,
Condition 1 can be used as a primary check, while Condition 2 must
be fulfilled by a decomposable 2-qubit gate.

\textbf{Example:} Here, as an example, we can consider the case of the
controlled-unitary gate given as $\left[\begin{array}{cc}
\mathbb{I} & 0\\
0 & U
\end{array}\right],$ where in place of $U$, we can use the general structure of unitary
matrix given in Eq. (\ref{eq:finalU}), using which we obtain

\begin{equation}
{\rm CU}=\left[\begin{array}{cccc}
1 & 0 & 0 & 0\\
0 & 1 & 0 & 0\\
0 & 0 & a\exp\left(i\phi\right) & b\exp\left(i\phi\right)\\
0 & 0 & -b^{*}\exp\left(i\phi\right) & a^{*}\exp\left(i\phi\right)
\end{array}\right].\label{eq:controlled-U}
\end{equation}
We can easily observe from Eq. (\ref{eq:controlled-U}) that the diagonal
elements do not have same modulus values so the gate is a genuine
2-qubit gate. Hence, we can easily observe that Condition
1 is violated, which gives the inseparability of the gate. 

Here, we can also check that violation of Condition 1 makes sure Condition
2 is also not satisfied. Specifically, when we attempt to write this
two qubit gate in the form of matrix $A$ of Eq. (\ref{eq:arbitrary2qubit}),
we will have
\begin{equation}
{\rm CU}=\exp\left(\frac{i\phi}{2}\right)\left[\begin{array}{cccc}
\exp\left(-\frac{i\phi}{2}\right) & 0 & 0 & 0\\
0 & \exp\left(-\frac{i\phi}{2}\right) & 0 & 0\\
0 & 0 & a\exp\left(\frac{i\phi}{2}\right) & b\exp\left(\frac{i\phi}{2}\right)\\
0 & 0 & -b^{*}\exp\left(\frac{i\phi}{2}\right) & a^{*}\exp\left(\frac{i\phi}{2}\right)
\end{array}\right],\label{eq:controlled-u2}
\end{equation}
which certainly fails Tests 1-3. Therefore, it concludes that the
controlled-unitary gate cannot be decomposed into two single qubit
gates. However, a special case, i.e., $a=1$ and $b=0$ becomes a
decomposable operation.

\section{Two single qubit gates from one tensor product gate\label{sec:Two-single-qubit}}

Once we have ensured that a 2-qubit gate is decomposable into two
single qubit gates, the important task in our hands would be to obtain
those two single qubit gates. In this section, we formulate a technique
to do so. From Eq. (\ref{eq:u1tensoru2}), if we consider only the
elements of the first two rows and two columns or more precisely the
first block of four elements, then we obtained 
\[
\left|\begin{array}{cc}
u_{1}v_{1} & u_{1}v_{2}\\
-u_{1}v_{2}^{*} & u_{1}v_{1}^{*}
\end{array}\right|=u_{1}^{2}\left(\left|v_{1}\right|^{2}+\left|v_{2}\right|^{2}\right)=u_{1}^{2}.
\]
Similarly, we can obtain this form for the remaining three blocks
of the matrix in (\ref{eq:u1tensoru2}) with four elements each. Specifically,
those can be written as 
\[
\left|\begin{array}{cc}
u_{2}v_{1} & u_{2}v_{2}\\
-u_{2}v_{2}^{*} & u_{2}v_{1}^{*}
\end{array}\right|=u_{2}^{2}\left(\left|v_{1}\right|^{2}+\left|v_{2}\right|^{2}\right)=u_{2}^{2};
\]
\[
\left|\begin{array}{cc}
u_{1}^{*}v_{1} & u_{1}^{*}v_{2}\\
-u_{1}^{*}v_{2}^{*} & u_{1}^{*}v_{1}^{*}
\end{array}\right|=u_{1}^{*2}\left(\left|v_{1}\right|^{2}+\left|v_{2}\right|^{2}\right)=u_{1}^{*2};
\]
\[
\left|\begin{array}{cc}
-u_{2}^{*}v_{1} & -u_{2}^{*}v_{2}\\
u_{2}^{*}v_{2}^{*} & -u_{2}^{*}v_{1}^{*}
\end{array}\right|=u_{2}^{*2}\left(\left|v_{1}\right|^{2}+\left|v_{2}\right|^{2}\right)=u_{2}^{*2}.
\]
Further, after comparing corresponding terms with the elements of
$A$, it can be easily obtained that 
\[
u_{1}^{2}=\left|\begin{array}{cc}
a & b\\
e & f
\end{array}\right|=af-be,
\]
\[
u_{2}^{2}=\left|\begin{array}{cc}
c & d\\
g & h
\end{array}\right|=ch-gd,
\]
\[
u_{1}^{*2}=\left|\begin{array}{cc}
k & l\\
o & p
\end{array}\right|=kp-lo,
\]
\[
u_{2}^{*2}=\left|\begin{array}{cc}
i & j\\
m & n
\end{array}\right|=in-jm,
\]
and
\[
v_{1}^{2}=\left|\begin{array}{cc}
a & c\\
i & k
\end{array}\right|=ak-ci,
\]
\[
v_{2}^{2}=\left|\begin{array}{cc}
b & d\\
j & l
\end{array}\right|=bl-jd,
\]
\[
v_{1}^{*2}=\left|\begin{array}{cc}
f & h\\
n & p
\end{array}\right|=fp-hn,
\]
\[
v_{2}^{*2}=\left|\begin{array}{cc}
e & g\\
m & o
\end{array}\right|=eo-gm.
\]
Using these relations we can formulate four conditions for decomposability,
which can be listed as follows\begin{subequations}
\begin{equation}
\left|\begin{array}{cc}
a & b\\
e & f
\end{array}\right|=\left(\left|\begin{array}{cc}
k & l\\
o & p
\end{array}\right|\right)^{*},\label{eq:C1}
\end{equation}
 
\begin{equation}
\left|\begin{array}{cc}
c & d\\
g & h
\end{array}\right|=\left(\left|\begin{array}{cc}
i & j\\
m & n
\end{array}\right|\right)^{*},\label{eq:C2}
\end{equation}
 
\begin{equation}
\left|\begin{array}{cc}
a & c\\
i & k
\end{array}\right|=\left(\left|\begin{array}{cc}
f & h\\
n & p
\end{array}\right|\right)^{*},\label{eq:C3}
\end{equation}
and 
\begin{equation}
\left|\begin{array}{cc}
b & d\\
j & l
\end{array}\right|=\left(\left|\begin{array}{cc}
e & g\\
m & o
\end{array}\right|\right)^{*}.\label{eq:C4}
\end{equation}
\end{subequations}These set of conditions do not ensure the non-decomposibility
of a 2-qubit gate. For example, one may consider a 2-qubit gate $\left[\begin{array}{cccc}
1 & 0 & 0 & 0\\
0 & 1 & 0 & 0\\
0 & 0 & \exp\left(i\phi\right) & 0\\
0 & 0 & 0 & \exp\left(-i\phi\right)
\end{array}\right]$, which satisfies these set of conditions, but Test 2 (in Condition
2) fails concluding in the non-decomposibility of this gate. Therefore,
we propose to use these results to obtain two single qubit gates once
Conditions 1-2 mentioned in the previous section are fulfilled.

Therefore, to write a two qubit gate as a tensor product of two single
qubit gates--if it is not a genuine two qubit gate--we can follow
this prescription and obtain the two single qubits as 

\[
U_{1}=\exp\left(i\phi_{1}\right)\left[\begin{array}{cc}
\sqrt{af-be} & \sqrt{ch-gd}\\
-\sqrt{in-jm} & \sqrt{kp-lo}
\end{array}\right]
\]
 and 
\[
U_{2}=\exp\left(-i\phi_{1}\right)\left[\begin{array}{cc}
\sqrt{ak-ci} & \sqrt{bl-jd}\\
-\sqrt{eo-gm} & \sqrt{fp-hn}
\end{array}\right].
\]
 Note that all the terms in both these unitary operations contain
square-root, which can be exploited to put a negative sign before
one of the terms in both the gates. As mentioned beforehand these
provide equivalent operations. In fact, even in this case, we obtain
a family of $U_{i}$s such that $U_{2}=\exp\left(-2i\phi_{1}\right)U_{1}$,
which are unitary in themselves, giving the same gate on their tensor
product. In what follows, we will show an example of a 2-qubit operation
in optical implementation.

\section{Applications\label{sec:Applications}}

In this section, we aim to illustrate the possible applications of
the results obtained in this work through some particular examples.
To begin with we consider an example that shows that the decomposition
of a polarization-dependent beam-splitter (PDBS) is not possible,
but a polarization-independent beam-splitter (PIDBS) can be decomposed.
Further, we have shown with the help of an explicit example that the
decomposability tests developed in the present work can be used to
reduce the quantum cost and gate counts of a quantum circuit.

\section*{Application 1: A physical example}

As an application of the present scheme, we would like to consider
the unitary operation corresponding to a PDBS \cite{PDBS}, i.e.,
\begin{equation}
{\rm PDBS}=\left(\begin{array}{cccc}
t_{aH} & ir_{bH} & 0 & 0\\
ir_{aH} & t_{bH} & 0 & 0\\
0 & 0 & t_{aV} & ir_{bV}\\
0 & 0 & ir_{aV} & t_{bV}
\end{array}\right).\label{eq:PDBS}
\end{equation}
Here, $t_{H}$ ($t_{V}$) and $r_{H}$ ($r_{V}$) are the transmission
and reflection coefficients of the PDBS for an incident photon in
horizontal (vertical) polarization state, respectively. Also, $a$
and $b$ mentioned in the subscript correspond to the two input ports
of the PDBS. Further, a phase change of $\frac{\pi}{2}$ is associated
with the reflected mode. PDBS is frequently used in implementing the
schemes of quantum communication and computation, and it may be viewed
as the most general type of BS. 

The PDBS can be checked with Conditions 1 and 2 to ensure that the
PDBS is a unitary operation that corresponds to a genuine 2-qubit
gate. In fact, this optical element is often used to generate entanglement.
Now, consider a special case- an usual beam-splitter which is independent
of polarization \cite{PDBS}, i.e., $t_{iH}=t_{iV}=t$ and $r_{iH}=r_{iV}=r$,
we obtain
\begin{equation}
{\rm PIDBS}=\left(\begin{array}{cccc}
t & ir & 0 & 0\\
ir & t & 0 & 0\\
0 & 0 & t & ir\\
0 & 0 & ir & t
\end{array}\right).\label{eq:PBS}
\end{equation}
It is important to note here that PIDBS satisfies all the requirements
(Tests 1-5) of a 2-qubit gate composed of two single qubit unitary
operations. One can also reconstruct the single qubit unitary operations
as follows

\[
U_{1}=\exp\left(i\phi_{1}\right)\left[\begin{array}{cc}
\sqrt{\left|t\right|^{2}+\left|r\right|^{2}} & 0\\
0 & \sqrt{\left|t\right|^{2}+\left|r\right|^{2}}
\end{array}\right]=\exp\left(i\phi_{1}\right)\left[\begin{array}{cc}
1 & 0\\
0 & 1
\end{array}\right]
\]
 and 
\[
U_{2}=\exp\left(-i\phi_{1}\right)\left[\begin{array}{cc}
t & ir\\
ir & t
\end{array}\right].
\]
The relevance of the unitary operations can be understood from the
input state of PDBS being $\left[\begin{array}{cccc}
a_{H} & b_{H} & a_{V} & b_{V}\end{array}\right]^{T}$, which can be viewed as $\left[\begin{array}{cc}
H & V\end{array}\right]^{T}\otimes\left[\begin{array}{cc}
a & b\end{array}\right]^{T}$, where polarization of the photon is an independent state acted upon
by an Identity and the evolution of the spatial modes of input photons
may be determined by a standard BS unitary operation ($U_{2}$ here).

\section*{Application 2: Optimization of quantum circuits}

\begin{figure}
\begin{centering}
\includegraphics[scale=0.67]{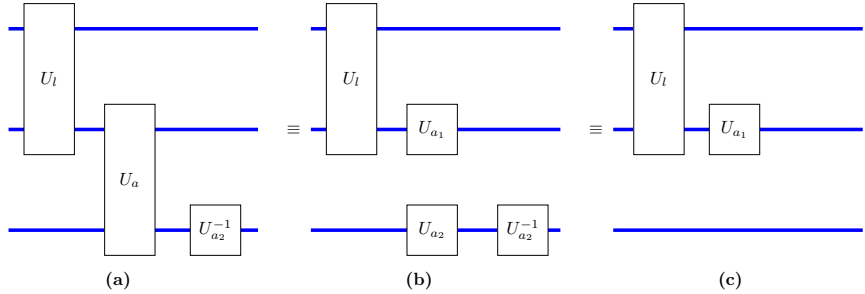}
\par\end{centering}

\protect\caption{\label{fig:cost}Three equivalent circuits are shown. The 2-qubit
gate $U_{l}$ ($U_{a}$) is assumed to be non-decomposable (decomposable)
and the decomposability of $U_{a}$ as $U_{a}=U_{a_{1}}\otimes U_{a_{2}}$
is used here to illustrate the role of decomposability test in reduction
of gate count, quantum cost and width of the circuit. It can be seen
that all these parameters are reduced in (c) with respect to (a). }
\end{figure}

The present result can be found useful in reducing quantum cost and
gate counts (circuit cost) of a given quantum circuit by exploiting
decomposability of 2-qubit gates, wherever possible. The minimum number
of elementary gates (1 qubit and 2-qubit gates) required to accomplish
a specific task is known as quantum cost \cite{rev1,quantumcost1}.
In fact, the use of single qubit gates in a qubit line that contains
one of the qubit of a 2-qubit gate does not increase the quantum cost
as the single qubit can be absorbed in the 2-qubit gate by creating
a new 1-qubit gate. Look at Fig. \ref{fig:cost} (a), it contains
two 2-qubit gates and a single-qubit gate. So its gate count or circuit
cost is three if we remain within the given gate library (before optimization).
However, the linear quantum cost (quantum cost obtained without circuit
optimization) would be two as for the determination of quantum cost
no restriction on gate library is imposed and one can always construct
a new 2-bit gate $U_{a}^{\prime}=U_{a}\left(\mathbb{I}_{2}\otimes U_{a_{2}}^{-1}\right)$.
This shows that single qubit gates usually do not contribute to the
quantum cost. Now, we assume that we apply the results of the previous
section and test the decomposability of both the 2-qubit gates present
in the circuit shown in Fig. \ref{fig:cost} (a), and our analysis
revealed that the leftmost 2-qubit gate $(U_{l})$ is a genuine 2-qubit
gate, whereas $U_{a}$ can be decomposed as $U_{a}=U_{a_{1}}\otimes U_{a_{2}}$.
The circuit after decomposition is shown in Fig. \ref{fig:cost} (b).
Clearly, $U_{a_{2}}U_{a_{2}}^{-1}=\mathbb{I}_{2},$ and we may remove
these gates and thus optimize the circuit as shown in Fig. \ref{fig:cost}
(c), which has a reduced circuit cost of two, and absorbing the single
qubit gate in the 2-qubit gate we would obtain the nonlinear quantum
cost (quantum cost obtained after optimization) as 1. Further, since
there is no operation in the third line, we can say that the width
of the quantum circuit is also reduced from three to two. Thus, this
simple example clearly illustrates the utility of the present work
(decomposability test designed here) in the optimization of quantum
cost, circuit cost and circuit width. In a similar manner, one can
show that this technique will be of help in optimizing quantum circuits
using template matching \cite{template-matching} and other methods
described in \cite{book,simplification} and references therein.

\section{Conclusion\label{sec:Conclusion}}

A general mathematical structure of single qubit gates is obtained
by exploiting their unitary nature. As the unitary nature of operations
ensures reversibility, the obtained result is applicable to classical
reversible gates, too. Further, unitarity of the operations only ensures
the existence of an operation (not necessarily the same) which can
transform the output state back into the input state. In other words,
unitarity only ensures reversibility, does not ensure
self reversibility. It is known that some of the quantum gates are
self-inverse, whereas the others are not. Keeping this in mind, the
mathematical form of the self-inverse unitary operations is also obtained
from the general structure of the single qubit gates. The restrictive
conditions obtained here, indicates that there are only a few self-inverse
single qubit quantum gates. 

The formulated general structure is subsequently used to obtain a
set of conditions using which one can easily test whether a given
arbitrary 2-qubit gate is non-decomposable or decomposable into two
single qubit gates. Once it is established that a given 2-qubit gate
is decomposable, the important task would be to obtain the decomposed
single qubit gates. Here, we have not only proposed an easy prescription
for reconstructing the decomposed single qubit operations, we have
also provided an interesting example related to the optical realization
of a quantum gate. Specifically, we have used a genuine entangled
gate for 2-qubits as PDBS, which is used frequently for entanglement
generation in quantum optical implementations. Subsequently, we have
shown mathematically that PDBS is not decomposable, but a polarization-independent
counterpart of it becomes decomposable. 

Finally, we have shown applications of the present results in the
optimization of quantum circuits. Specifically, we have shown that
when a 2-qubit quantum gate is decomposable, employing its decomposed
single qubit counterparts can be used to reduce quantum cost, gate
counts and circuit width. We conclude the paper with a hope that the
present results would find wider scale applications in the synthesis
and optimization of reversible and quantum circuits, and some of the
circuits designed and/or optimized using the decomposability test
developed here will be tested using IBM Quantum Experience, a quantum
computer places in cloud \cite{IBM} or other quantum hardwares capable
of performing the task.

\textbf{Acknowledgment:} KT and AP thank Defense Research \& Development
Organization (DRDO), India for the support provided through the project
number ERIP/ER/1403163/M/01/1603.

\end{document}